\documentclass{webofc}

\usepackage[varg]{txfonts}   
\usepackage{hyperref}
\usepackage{url}
\hypersetup{colorlinks=true,citecolor=blue,urlcolor=blue,linkcolor=blue}
\newcommand{\Pom}{\mathbb{P}}
\newcommand{\Reg}{\mathbb{R}}
\newcommand{\Ode}{\mathbb{O}}

\usepackage{braket}

\begin{document}
\title{Exclusive production of $\pi^+ \pi^-$ pairs in diffractive $\gamma p$ and \\in $pp$ collisions within the tensor-pomeron approach}
%
%

\author{\firstname{Piotr} \lastname{Lebiedowicz}\inst{1}\fnsep\thanks{\email{Piotr.Lebiedowicz@ifj.edu.pl}} \and
        \firstname{Otto} \lastname{Nachtmann}\inst{2}\fnsep\thanks{\email{O.Nachtmann@thphys.uni-heidelberg.de}} \and
        \firstname{Antoni} \lastname{Szczurek}\inst{1, 3}\fnsep\thanks{\email{Antoni.Szczurek@ifj.edu.pl}}
}

\institute{Institute of Nuclear Physics Polish Academy of Sciences, \\
Radzikowskiego 152, PL-31342 Krak\'ow, Poland
\and
           Institut f\"ur Theoretische Physik, Universit\"at Heidelberg, \\
Philosophenweg 16, D-69120 Heidelberg, Germany
\and
           Institute of Physics, Faculty of Exact and Technical Sciences, University of Rzesz\'ow,\\
Pigonia 1, PL-35310 Rzesz\'ow, Poland
          }

\abstract{We discuss exclusive production of $\pi^+ \pi^-$ pairs
in diffractive $\gamma p$ and in $pp$ collisions at high energies, 
considering resonant ($\rho(770)$, $\omega$, $f_2(1270)$) 
and non-resonant (Drell-S\"oding) 
contributions within the tensor-pomeron approach.
For the $\gamma p \to \pi^+ \pi^- p$ reaction, 
the model describes well the H1 data in the region $M_{\pi \pi} \lesssim 1.2$ GeV.
We discuss the important role of the Drell-S\"oding mechanism 
in shaping the resonance line.
We also predict differential cross sections for
the $pp \to pp \pi^+ \pi^-$ reaction at $\sqrt{s} = 13$ TeV
where at least one proton emits a virtual photon.
These findings are relevant for the central exclusive production 
of $\pi^+ \pi^-$ pairs 
in the context of ALICE, ATLAS, CMS, and LHCb measurements 
in hadron-hadron collisions at the LHC, 
especially under experimental selections using 
only rapidity-gap conditions.
Additionally, our results are also applicable to $\pi^+ \pi^-$ production
in ultraperipheral $p$A/AA collisions 
at the LHC 
and to photo- and electroproduction of $\pi^+ \pi^-$ 
at future electron-proton/ion colliders (EIC, LHeC).}


%
\maketitle
\section{Introduction}
\label{intro}

In this contribution, we discuss the production of $\pi^+ \pi^-$ pairs 
in photon-proton collisions,
and the central exclusive production (CEP) 
of such pairs in proton-proton collisions,
\begin{eqnarray}
&&\gamma^{(*)}  p \to \pi^+ \pi^- p \,, 
\label{1.1a}\\
&&p p \to p p \pi^+ \pi^- \,.
\label{1.1b}
\end{eqnarray}
Here, $\gamma^{(*)}$ denotes a real or slightly virtual photon.
Our focus is on $\pi^{+}\pi^{-}$ invariant
masses in the region of the $\rho^{0}(770)$ resonance and below, $M_{\pi \pi} \lesssim 1$~GeV.
Compared to the $\rho^{0}$ line shape measured in $e^{+}e^{-}$ annihilation,
there is a distinct skewing of the $\rho^{0}$ shape observed in these hadronic reactions.
This skewing is attributed to the interference
of the decay $\rho^{0} \to \pi^{+}\pi^{-}$
with the non-resonant production of $\pi^{+}\pi^{-}$,
known as the Drell-S\"oding mechanism \cite{Drell:1960zz,Soding:1965nh}.
In practice, the field-theoretic calculation of this term is a challenging problem,
largely due to the strict requirements of gauge invariance;
see e.g. \cite{Pumplin:1970kp,Szczurek:2004xe,Bolz:2014mya,Lebiedowicz:2014bea,JointPhysicsAnalysisCenter:2024qld,Lebiedowicz:2025xob,Lebiedowicz:2026ixn}.

We employ the framework of the tensor-pomeron model \cite{Ewerz:2013kda} which
has been constructed to describe soft high-energy hadronic reactions.
In this approach, the pomeron ($\Pom$) and the charge-conjugation $C = +1$ reggeons 
($f_{2 \Reg}$, $a_{2 \Reg}$)
are treated as effective rank-two symmetric tensor exchanges,
while the odderon ($\Ode$) and the $C = -1$ reggeons 
($\rho_{\Reg}$, $\omega_{\Reg}$)
are treated as effective vector exchanges.
The tensor-pomeron model has been successfully applied to various central exclusive production (CEP) 
reactions in proton-proton collisions 
(see, e.g., \cite{Lebiedowicz:2016ioh,Lebiedowicz:2025num}).

In \cite{Bolz:2014mya}, the photoproduction of a $\pi^{+} \pi^{-}$ pair (\ref{1.1a})
was studied theoretically using this tensor-pomeron approach \cite{Ewerz:2013kda}.
For the calculation of the Drell-S\"oding term,
an effective gauge-invariant coupling Lagrangian was introduced
to describe the $\gamma \pi \pi$, $\gamma \gamma \pi \pi$, 
$\Pom \pi \pi$, and $\Pom \gamma \pi \pi$ vertices.
In the various diagrams arising in the calculation 
of the Drell-S\"oding (DS) term, a common energy variable
was assumed in the respective Regge factors.
Although this prescription achieved a skewing of the $\rho^{0}$ shape,
the effect was not large enough when compared to experimental data;
see for instance \cite{H1:2020lzc,Bolz_Meson2021}.
The same ansatz was used by us in \cite{Lebiedowicz:2014bea}
for the CEP of $\pi^{+} \pi^{-}$ in the reaction (\ref{1.1b}).

\begin{figure}[h]
\centering
\includegraphics[width=4.7cm,clip]{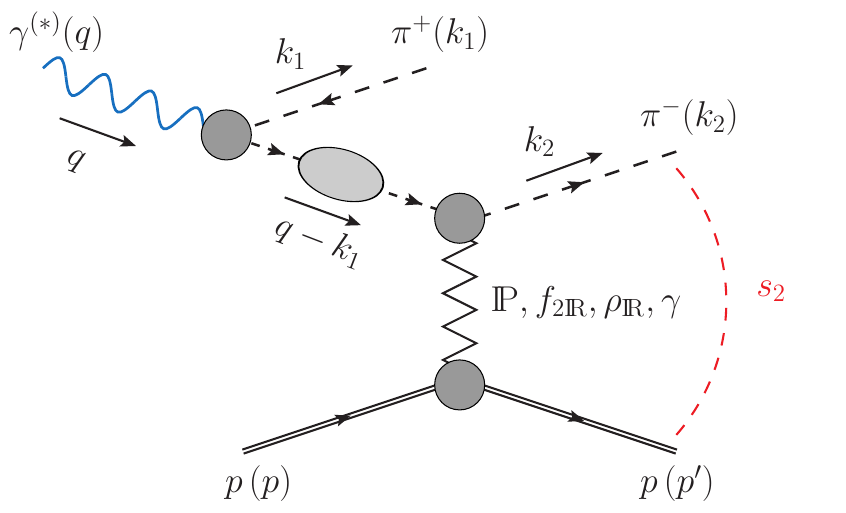}
\includegraphics[width=4.7cm,clip]{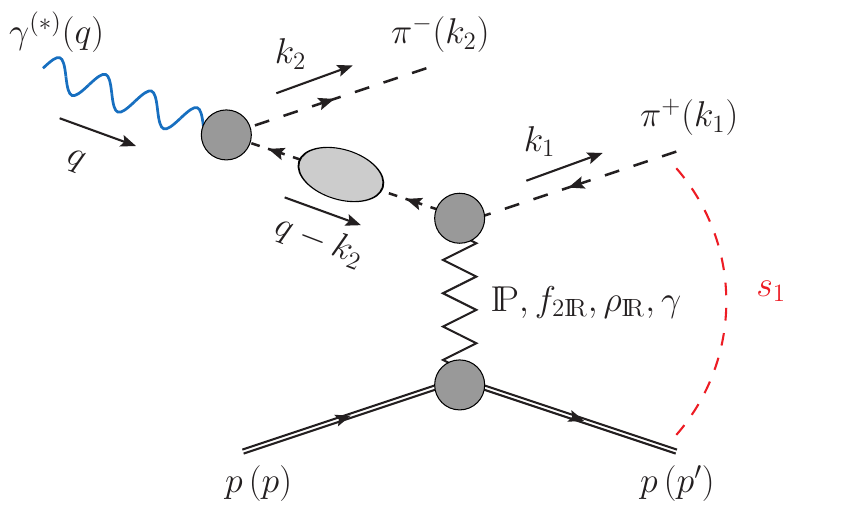}
\includegraphics[width=3.4cm,clip]{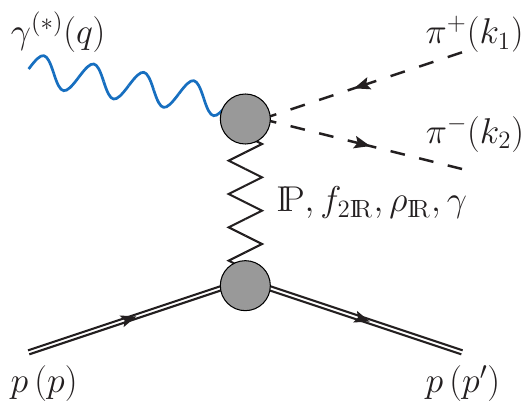}
\caption{Diagrams for non-resonant $\pi^{+}\pi^{-}$ production 
in the reaction $\gamma^{(*)}  p \to \pi^+ \pi^- p$ with pomeron ($\Pom$),
reggeons ($f_{2 \Reg}$, $\rho_{\Reg}$), and photon ($\gamma$) exchange.
The blobs denote the full vertices and the complete pion propagators.}
\label{fig:1}
\end{figure}

Recently, in \cite{Lebiedowicz:2025xob}, we presented an improved calculation 
of the Drell-S\"oding term, representing non-resonant $\pi^{+}\pi^{-}$ production 
(see figure~\ref{fig:1}).
Furthermore, in \cite{Lebiedowicz:2026ixn} we updated the DS amplitudes
by explicitly taking into account pion off-shell effects
in the $\Pom \pi \pi$, $f_{2 \Reg} \pi \pi$, and $\rho_{\Reg} \pi \pi$ vertices.
There, we have presented the results of our model for the real photoproduction
of $\pi^{+}\pi^{-}$ pairs and compared them with data measured by
the H1 \cite{H1:2020lzc} and ZEUS \cite{ZEUS:1997rof} Collaborations.
%

\section{Sketch of the formalism}
\label{sec-2}

The presentation is based on \cite{Lebiedowicz:2025xob},
where all details and many more results 
for the reactions (\ref{1.1a}) and (\ref{1.1b}) 
calculated within the tensor-pomeron approach can be found.
To accurately reproduce the spectral $\rho$ line shape, 
the $\gamma^{(*)} p \to \pi^+ \pi^- p$ formalism explicitly 
accounts for pion off-shell effects in the $\Pom/\Reg \pi \pi$ vertices by introducing the respective form factor functions, 
as explained in \cite{Lebiedowicz:2026ixn}

\subsection{The $\gamma^{(*)} p \to \pi^+ \pi^- p$ reaction}
\label{sec-2a}

\begin{figure}[h]
\centering
(a)\includegraphics[width=4.7cm,clip]{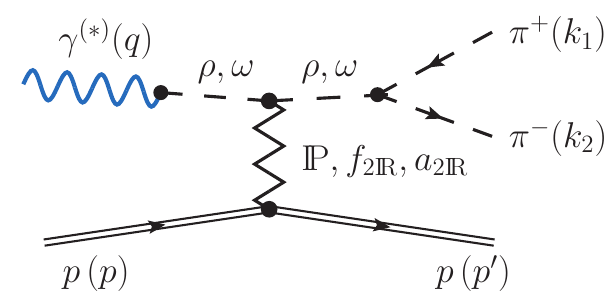}
(b)\includegraphics[width=4.7cm,clip]{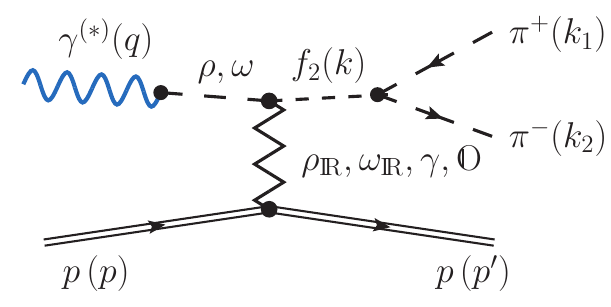}
\caption{Diagrams for resonant $\pi^{+}\pi^{-}$ production 
in the reaction $\gamma^{(*)}  p \to \pi^+ \pi^- p$:
(a) production of the vector-mesons $\rho$ and $\omega$
which subsequently decay into $\pi^{+}\pi^{-}$;
(b) production of the $f_{2}(1270)$ meson.}
\label{fig:2}
\end{figure}

We consider the reaction (\ref{1.1a}) 
\begin{eqnarray}
\gamma^{(*)}(q, \mu) + p(p, \mathfrak{s}) \to \pi^{+}(k_{1}) + \pi^{-}(k_{2}) + p(p',\mathfrak{s'})\,.
\label{2.1}
\end{eqnarray}
We denote by $k_{1}$, $k_{2}$, $p$, $p'$, and $q$
the four-momenta of the involved particles,
by $\mu$ the vector index of $\gamma^{(*)}$,
and by $\mathfrak{s}$ and $\mathfrak{s'}$ the spin indices of the incoming and outgoing protons, respectively.
The matrix element for the reaction (\ref{2.1}) 
with a real photon of the polarization vector $\epsilon$
is
\begin{eqnarray}
\braket{\pi^{+}(k_{1}),\pi^{-}(k_{2}),p(p',\mathfrak{s'})|{\cal T}|
\gamma(q, \epsilon),p(p,\mathfrak{s})} = \epsilon^{\mu}
{\cal M}_{\mu, \mathfrak{s'},\mathfrak{s}}(k_{1},k_{2},p',q,p)\,.
\label{2.2}
\end{eqnarray}
The complete amplitude is given by
\begin{eqnarray}
{\cal M}_{\mu, \mathfrak{s'},\mathfrak{s}}
=
{\cal M}^{(\rm DS)}_{\mu, \mathfrak{s'},\mathfrak{s}}
|_{\Pom + f_{2 \Reg} + \rho_{\Reg} + \gamma}
+
{\cal M}^{(\rm res)}_{\mu, \mathfrak{s'},\mathfrak{s}}
|_{\Pom + f_{2 \Reg} + a_{2 \Reg}}
+
{\cal M}^{(f_{2})}_{\mu, \mathfrak{s'},\mathfrak{s}}
|_{\rho_{\Reg} + \omega_{\Reg} + \Ode + \gamma}
\,.
\label{2.2b}
\end{eqnarray}
Here, the total amplitude is expressed as the sum of the non-resonant Drell-Söding term (${\cal M}^{(\rm DS)}_{\mu, \mathfrak{s'},\mathfrak{s}}$), resonant vector-meson production (${\cal M}^{(\rm res)}_{\mu, \mathfrak{s'},\mathfrak{s}}$) driven by C-even exchanges, and tensor $f_{2}(1270)$ meson production (${\cal M}^{(f_{2})}_{\mu, \mathfrak{s'},\mathfrak{s}}$) mediated by C-odd exchanges. 
The diagrams for the DS term with the available exchanges 
are shown in figure~\ref{fig:1}.
The corresponding diagrams for resonant $\pi^{+}\pi^{-}$ production 
are shown in figure~\ref{fig:2}.

In the following, we consider only the leading pomeron-exchange contribution for the DS term.
We write
\begin{eqnarray}
{\cal M}^{(\rm DS)}_{\mu, \mathfrak{s'},\mathfrak{s}}|_{\Pom} 
= {\cal M}_{\mu, \mathfrak{s'},\mathfrak{s}}^{(a)}|_{\Pom} + 
{\cal M}_{\mu, \mathfrak{s'},\mathfrak{s}}^{(b)}|_{\Pom} 
+
{\cal M}_{\mu, \mathfrak{s'},\mathfrak{s}}^{(c)}|_{\Pom}\,. \qquad
\label{2.3}
\end{eqnarray}
With the expressions for the vertex functions, propagators,
and the pion-proton amplitudes
${\cal M}_{\mathfrak{s'},\mathfrak{s}}^{(0, \,a)}|_{\Pom}$
and ${\cal M}_{\mathfrak{s'},\mathfrak{s}}^{(0, \,b)}|_{\Pom}$
from Appendix~A of \cite{Lebiedowicz:2025xob},
we obtain
\begin{align}
{\cal M}_{\mu, \mathfrak{s'},\mathfrak{s}}^{(a)}(k_{1},k_{2},p',q,p) |_{\Pom} 
&=
e \widehat{\Gamma}_{\mu}^{(\gamma \pi \pi)}(k_{1},k_{1}-q) 
\Delta_{F}[(k_{1}-q)^{2}]
{\cal M}_{\mathfrak{s'},\mathfrak{s}}^{(0, \,a)}|_{\Pom}\,,\nonumber \\
{\cal M}_{\mathfrak{s'},\mathfrak{s}}^{(0, \,a)}|_{\Pom} \equiv
{\cal M}_{\mathfrak{s'},\mathfrak{s}}^{(0, \,a)}(k_{2},p',q-k_{1},p)|_{\Pom} &=
{\cal M}_{\mathfrak{s'},\mathfrak{s}}^{(\pi^{-})}(k_{2},p',q-k_{1},p)|_{\Pom}\,, \qquad
\label{2.4}
\end{align}
\begin{align}
{\cal M}_{\mu, \mathfrak{s'},\mathfrak{s}}^{(b)}(k_{1},k_{2},p',q,p) |_{\Pom} 
&=
-e \widehat{\Gamma}_{\mu}^{(\gamma \pi \pi)}(k_{2},k_{2}-q) 
\Delta_{F}[(k_{2}-q)^{2}]
{\cal M}_{\mathfrak{s'},\mathfrak{s}}^{(0, \,b)}|_{\Pom}\,,\nonumber \\
{\cal M}_{\mathfrak{s'},\mathfrak{s}}^{(0, \,b)}|_{\Pom} \equiv
{\cal M}_{\mathfrak{s'},\mathfrak{s}}^{(0, \,b)}(k_{1},p',q-k_{2},p)|_{\Pom} &=
{\cal M}_{\mathfrak{s'},\mathfrak{s}}^{(\pi^{+})}(k_{1},p',q-k_{2},p)|_{\Pom}\,. \qquad
\label{2.5}
\end{align}
Gauge invariance requires that
\begin{eqnarray}
q^{\mu} {\cal M}^{(\rm DS)}_{\mu, \mathfrak{s'},\mathfrak{s}}|_{\Pom} = 0\,.
\label{2.6}
\end{eqnarray}
Using the generalized Ward identity for the pion field 
[see equation~(B39) of \cite{Lebiedowicz:2023mlz}]
\begin{eqnarray}
(k'-k)^{\mu} \, \widehat{\Gamma}_{\mu}^{(\gamma \pi\pi)}(k',k) =
\Delta_{F}^{-1}(k'^{2}) - \Delta_{F}^{-1}(k^{2})\,,
\label{A9}
\end{eqnarray}
we find
\begin{eqnarray}
q^{\mu} {\cal M}_{\mu, \mathfrak{s'},\mathfrak{s}}^{(a)}|_{\Pom} 
= -e {\cal M}_{\mathfrak{s'},\mathfrak{s}}^{(0, \,a)}|_{\Pom}\,, \qquad 
q^{\mu} {\cal M}_{\mu, \mathfrak{s'},\mathfrak{s}}^{(b)}|_{\Pom} 
= e {\cal M}_{\mathfrak{s'},\mathfrak{s}}^{(0, \,b)}|_{\Pom}\,.
\label{2.7}
\end{eqnarray}
Therefore, we have
\begin{align}
q^{\mu} {\cal M}_{\mu, \mathfrak{s'},\mathfrak{s}}^{(c)}(k_{1},k_{2},p',q,p)|_{\Pom} 
= e 
\left[ 
{\cal M}_{\mathfrak{s'},\mathfrak{s}}^{(0, \,a)}(k_{2},p',q-k_{1},p)|_{\Pom}
-{\cal M}_{\mathfrak{s'},\mathfrak{s}}^{(0, \,b)}(k_{1},p',q-k_{2},p)|_{\Pom}
\right] \,. 
\label{2.8}
\end{align}

All kinematic relations for the amplitude (\ref{2.1}) are defined 
in Appendix~A of \cite{Lebiedowicz:2025xob}.
Let us recall some of them.
We have energy-momentum conservation
\begin{eqnarray}
q + p = k_{1} + k_{2} + p'
\label{A1}
\end{eqnarray}
and the following variables:
\begin{align}
s &= (q + p)^{2} = (k_{1} + k_{2} + p')^{2}\,,
\quad \sqrt{s} = W_{\gamma p}\,, 
\quad t = (p - p')^{2}\,, \quad
M_{\pi \pi}^{2} = (k_{1} + k_{2})^{2} \,, \nonumber \\
s_{1} &= (p' + k_{1})^{2} \,, \quad 
s_{2} =  (p' + k_{2})^{2} \,, \quad
u_{1} = (p - k_{1})^{2} \,, \quad
u_{2} = (p - k_{2})^{2} \,, \nonumber \\
\nu_{1} &= \frac{1}{4}(s_{1} - u_{1})\,, \quad
\nu_{2} = \frac{1}{4}(s_{2} - u_{2}) \,.
\label{A3}
\end{align}
We define
\begin{eqnarray}
\bar{\nu}^{2} = \frac{1}{2}(\nu_{1}^{2} + \nu_{2}^{2})\,, \quad
\varkappa = \frac{2(q,p+p')(p+p',k_{1}-k_{2})}{16 \bar{\nu}^{2}}\,, \quad
g(\lambda, \varkappa) = \frac{(1 - \varkappa)^{-\lambda} - 1}{\lambda \varkappa} \,.
\label{A5}
\end{eqnarray}
For diagrams~(a) and (b) of figure~\ref{fig:1}, we get for $\Pom$ exchange
\begin{align}
{\cal M}_{\mu, \mathfrak{s'},\mathfrak{s}}^{(a)}(k_{1},k_{2},p',q,p)|_{\Pom}
&=
e 
\left[
\frac{(2 k_{1} - q)_{\mu}}{- 2 k_{1} \cdot q + q^{2} + i \varepsilon}
F_{M}(q^{2}) - q_{\mu} \frac{1 - F_{M}(q^{2})}{q^{2}}
\right]
{\cal M}_{\mathfrak{s'},\mathfrak{s}}^{(0, \,a)}|_{\Pom}\,, \nonumber \\ 
{\cal M}_{\mathfrak{s'},\mathfrak{s}}^{(0, \,a)}|_{\Pom} 
&=
i{\cal F}_{\Pom \pi p}(2 \bar{\nu}, t)
\left[
1 + (2 - \alpha_{\Pom}(t)) \frac{\varkappa}{2} \,
g\left( \frac{2 - \alpha_{\Pom}(t)}{2}, \varkappa \right)
\right] \nonumber\\
& \times
\Big[
2(k_{2}-k_{1}+q)^{\nu}(k_{2}-k_{1}+q,p'+p)
\bar{u}_{\mathfrak{s'}}(p') \gamma_{\nu} u_{\mathfrak{s}}(p) 
\nonumber\\
& 
-(k_{2}-k_{1}+q)^{2} m_{p}
\bar{u}_{\mathfrak{s'}}(p') u_{\mathfrak{s}}(p)
\Big] \,,
\label{2.9}
\end{align}
\begin{align}
{\cal M}_{\mu, \mathfrak{s'},\mathfrak{s}}^{(b)}(k_{1},k_{2},p',q,p)|_{\Pom}
&= -e \left[
\frac{ (2 k_{2} - q)_{\mu}}{-2 k_{2} \cdot q + q^{2} + i \varepsilon}
F_{M}(q^{2}) - q_{\mu} \frac{1 - F_{M}(q^{2})}{q^{2}}
\right] {\cal M}_{\mathfrak{s'},\mathfrak{s}}^{(0,\,b)}|_{\Pom}\,, \nonumber \\
{\cal M}_{\mathfrak{s'},\mathfrak{s}}^{(0,\,b)}|_{\Pom}
&=
i{\cal F}_{\Pom \pi p}(2 \bar{\nu}, t)
\left[
1 - (2 - \alpha_{\Pom}(t)) \frac{\varkappa}{2} \,
g\left( \frac{2 - \alpha_{\Pom}(t)}{2}, -\varkappa \right)
\right] \nonumber\\
& \times
\Big[
2(k_{2}-k_{1}-q)^{\nu}(k_{2}-k_{1}-q,p'+p)
\bar{u}_{\mathfrak{s'}}(p') \gamma_{\nu} u_{\mathfrak{s}}(p)  
\nonumber \\
&-
(k_{2}-k_{1}-q)^{2} m_{p}
\bar{u}_{\mathfrak{s'}}(p') u_{\mathfrak{s}}(p)
\Big] \,;
\label{2.10}
\end{align}
see equations (2.12) and (2.13) of \cite{Lebiedowicz:2025xob}.
Here,
\begin{eqnarray}
F_{M}(q^{2}) = \frac{m_{0}^{2}}{m_{0}^{2}-q^{2}}\,, \quad 
m_{0}^{2} = 0.50~{\rm GeV}^{2}\,
\label{A21b}
\end{eqnarray}
is a simple representation of the pion electromagnetic 
form factor $F_{\rm em}^{(\pi)}(q^{2})$;
see equation~(3.34) of \cite{Ewerz:2013kda}.
Furthermore,
\begin{eqnarray}
{\cal F}_{\Pom \pi p}(2 \bar{\nu}, t) 
=
6 \,\beta_{\Pom \pi \pi}\, \beta_{\Pom NN}\,
F_{M}(t) \,
F_{1}(t) 
\frac{1}{8 \bar{\nu}}(-i \, 2 \bar{\nu}\, \alpha'_{\Pom})^{\alpha_{\Pom}(t)-1} \,,
\label{A32}
\end{eqnarray}
where $\beta_{\Pom \pi \pi} = 1.76$~GeV$^{-1}$ and
$\beta_{\Pom NN} = 1.87$~GeV$^{-1}$.
We use the electromagnetic Dirac form factor $F_{1}(t)$ of the proton
for the pomeron-proton coupling,
while for the pomeron-pion coupling, the same form 
is used as in equation~(\ref{A21b})
but with $m_{0}^{2} \to \Lambda^{2} = 0.75$~GeV$^{2}$
(see Appendix~B of \cite{Lebiedowicz:2025xob}).
The pomeron trajectory $\alpha_{\Pom}(t)$ is
assumed to have the standard form:
\begin{eqnarray}
\alpha_{\Pom}(t) = \alpha_{\Pom}(0)+\alpha'_{\Pom} t\,,
\quad 
\alpha_{\Pom}(0) = 1 + \epsilon_{\Pom} = 1.0808\,, 
\quad
\alpha'_{\Pom} = 0.25 \; \mathrm{GeV}^{-2}\,.
\label{pomtrajectory}
\end{eqnarray}

Using (\ref{2.9}) and (\ref{2.10}), 
one can write the right-hand side of (\ref{2.8}) 
in a way that 
is explicitly $\propto q^{\mu}$.
Then we drop $q^{\mu}$ on both sides and get the simplest solution 
for ${\cal M}_{\mu, \mathfrak{s'},\mathfrak{s}}^{(c)}$, 
the amplitude for the diagram (c) of figure~\ref{fig:1},
which is satisfactory from the QFT point of view.
See equation (2.23) of \cite{Lebiedowicz:2025xob}.

We treat the $C=+1$ $f_{2 \Reg}$-exchange term analogously 
to the pomeron-exchange term. 
The corresponding amplitudes have the same structure, but with the following replacements:
\begin{eqnarray}
{\cal F}_{\Pom \pi p}(2 \bar{\nu}, t) 
\to {\cal F}_{f_{2 \Reg} \pi p}(2 \bar{\nu}, t)\,, 
\qquad
\alpha_{\Pom}(t) 
\to \alpha_{f_{2 \Reg}}(t) \,;
\label{2.22}
\end{eqnarray}
see Sect.~2.2.3 of \cite{Lebiedowicz:2025xob}. 
The expressions for non-resonant $\pi^{+} \pi^{-}$ production
via $C = -1$ $\rho_{\Reg}$ and $\gamma$ exchanges,
${\cal M}^{(\rm DS)}_{\mu, \mathfrak{s'},\mathfrak{s}}|_{\rho_{\Reg}}$
and
${\cal M}^{(\rm DS)}_{\mu, \mathfrak{s'},\mathfrak{s}}|_{\gamma}$,
are given in Sect.~2.2.4 of \cite{Lebiedowicz:2025xob}. 
We stress that the $\gamma$-exchange term plays an important role 
at very small momentum transfer $|t|$.

It should be emphasized that in \cite{Lebiedowicz:2026ixn} 
we have generalized our DS model 
by allowing for off-shell effects in the $\Pom \pi \pi$,
$f_{2 \Reg} \pi \pi$, and $\rho_{\Reg} \pi \pi$ vertices 
occurring in figures~\ref{fig:1}(a) and (b).
We employ the off-shell pion form factors
\begin{eqnarray}
F_{\pi}(t_{\pi,i}) = \exp\left( \frac{t_{\pi,i} - m_{\pi}^2}{\Lambda_{\pi}^2} \right)\,, \quad i = 1, 2\,,
\label{form_factor_off}
\end{eqnarray} 
where $t_{\pi,i} = (q - k_{i})^{2}$
and the cutoff parameter is chosen as $\Lambda_{\pi} = 0.8$~GeV.
For a detailed discussion of the off-shell effects, 
we refer to \cite{Lebiedowicz:2026ixn}.
Note that there, the amplitudes $\mathcal{M}^{(a)}_{\mu}$ 
and $\mathcal{M}^{(b)}_{\mu}$ are generalized to new amplitudes 
$\mathcal{N}^{(a)}_{\mu}$ and $\mathcal{N}^{(b)}_{\mu}$ by 
incorporating the off-shell pion form factors. Consequently, 
the amplitude $\mathcal{M}^{(c)}_{\mu}$ is updated to 
$\mathcal{N}^{(c)}_{\mu}$, ensuring that the total amplitude 
satisfies $q^{\mu} \mathcal{N}_{\mu} = 0$.

Let us now briefly turn to the resonant di-pion production.
The amplitude for vector-meson production
denoted in (\ref{2.2b}) by
${\cal M}^{(\rm res)}_{\mu, \mathfrak{s'},\mathfrak{s}}
|_{\Pom + f_{2 \Reg} + a_{2 \Reg}}$
[see diagram (a) of figure~\ref{fig:2}]
for real photons is given in subsection 2.1 of \cite{Bolz:2014mya}.
In Sect.~2.2.1 of \cite{Lebiedowicz:2025xob},
we extended these formulas to take virtual photons into account as well.
We include the $\rho$--$\omega$ interference effect
in the final state via propagator mixing \cite{Melikhov:2003hs}
and the explicit $\omega \to \pi^{+} \pi^{-}$ decay.
In the initial state where the photon turns into a $\rho$ or $\omega$
meson we neglect $\rho$--$\omega$ mixing.
The amplitude for the production of the $f_{2}(1270)$ meson
[see figure~\ref{fig:2}(b)]
denoted in (\ref{2.2b}) by
${\cal M}^{(f_{2})}_{\mu, \mathfrak{s'},\mathfrak{s}}
|_{\rho_{\Reg} + \omega_{\Reg} + \Ode + \gamma}$
is discussed in detail in Sect.~2.2.2 of \cite{Lebiedowicz:2025xob}.

\subsection{The $pp \to pp \pi^+ \pi^-$ reaction}
\label{sec-2b}

\begin{figure}[h]
\centering
\includegraphics[width=4.cm,clip]{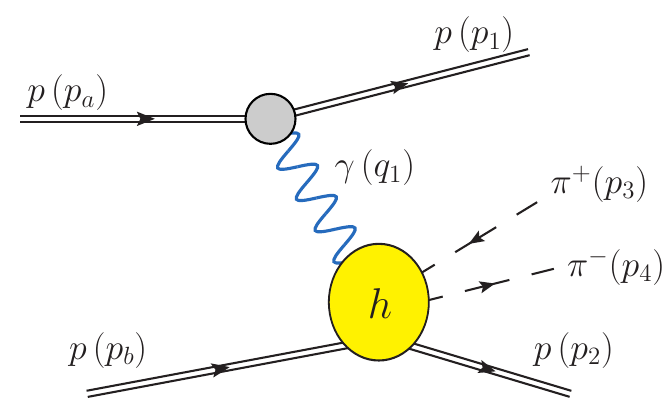}
\includegraphics[width=4.cm,clip]{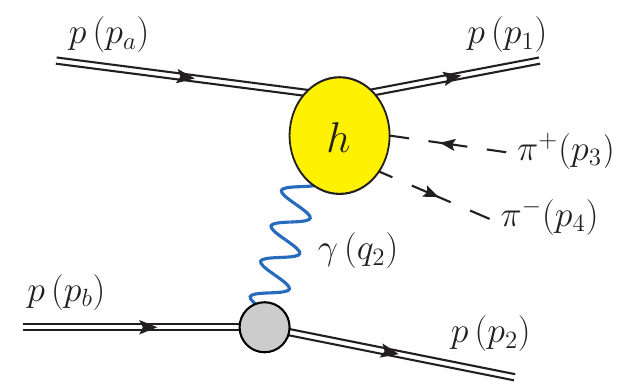}
\includegraphics[width=4.cm,clip]{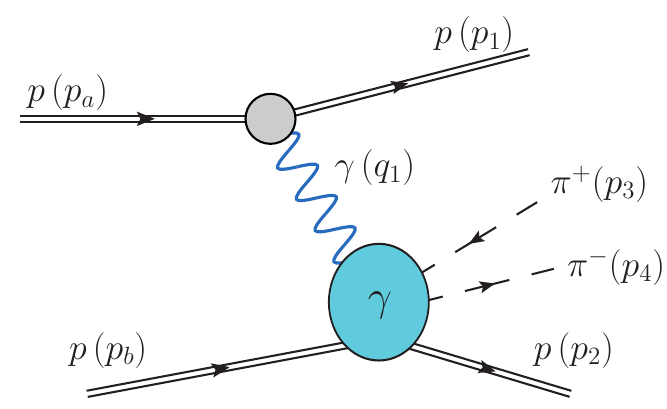}
\caption{Diagrams for the reaction (\ref{3.1}) in the diffractive regime
where we require photon exchange at least for one of the incoming protons.
The diagrams (a) and (b) describe the reactions where
one proton emits a photon and the other proton reacts
hadronically, as indicated by the blob with the $h$.
Here, $h$ denotes the hadronic exchange,
including both pomeron and reggeon contributions
depending on the specific production process.
Diagram (c) stands for the $\gamma \gamma$ fusion contribution
where both incoming protons emit a photon.}
\label{fig:3}
\end{figure}

Now we discuss the central exclusive production (CEP)
of $\pi^{+} \pi^{-}$ in proton-proton collisions,
\begin{eqnarray}
p(p_{a},\mathfrak{s}_{a}) + p(p_{b},\mathfrak{s}_{b}) \to
p(p_{1},\mathfrak{s}_{1}) + \pi^{+}(p_{3}) + \pi^{-}(p_{4}) + p(p_{2},\mathfrak{s}_{2}) \,,
\label{3.1}
\end{eqnarray}
where $p_{a,b}$, $p_{1,2}$ and 
$\mathfrak{s}_{a,b}$, $\mathfrak{s}_{1,2} \in \{1/2, -1/2\}$ 
denote the four-momenta and spin indices of the protons, 
and $p_{3,4}$ denote the four-momenta of the charged pions, respectively.
We shall study $\pi^{+}\pi^{-}$ production
where at least one of the protons emits a virtual photon;
see figure~\ref{fig:3}.
It should be noted that also purely hadronic diagrams will also contribute,
where only pomeron, odderon, and reggeon exchanges 
are involved in the production of the pion pair.
These latter reactions have been discussed in \cite{Lebiedowicz:2016ioh}.

The complete amplitude for (\ref{3.1}) is
\begin{eqnarray}
{\cal M}_{pp \to pp \pi^{+} \pi^{-}} =
{\cal M}^{(\gamma h)}_{pp \to pp \pi^{+} \pi^{-}} +
{\cal M}^{(h \gamma)}_{pp \to pp \pi^{+} \pi^{-}} +
{\cal M}^{(\gamma \gamma)}_{pp \to pp \pi^{+} \pi^{-}} \,.
\label{3.8} 
\end{eqnarray}
From diagram~(a) of figure~\ref{fig:3}, we have
\begin{eqnarray}
{\cal M}^{(\gamma h)}_{pp \to pp \pi^{+} \pi^{-}} =
\bar{u}_{\mathfrak{s}_{1}}(p_{1}) 
\Gamma^{(\gamma pp)\,\mu}(p_{1},p_{a}) 
u_{\mathfrak{s}_{a}}(p_{a}) \frac{1}{t_{1}} 
{\cal M}_{\mu, \mathfrak{s}_{2}, \mathfrak{s}_{b}}(p_{3},p_{4},p_{2},q_{1},p_{b})|_{h}\,.
\label{3.5} 
\end{eqnarray}
The photon-proton vertex function
$\Gamma^{(\gamma pp)\,\mu}$ is given in equations 
(3.26)--(3.32) of \cite{Ewerz:2013kda},
and ${\cal M}_{\mu, \mathfrak{s}_{2}, \mathfrak{s}_{b}}|_{h}$
is given by (\ref{2.2b}) with the exception of the $\gamma$ exchange. 
In the calculations, we use the high-energy approximation
\begin{eqnarray}
\bar{u}_{\mathfrak{s'}}(p') \,
\Gamma^{(\gamma pp)\,\mu} (p', p) \,u_{\mathfrak{s}}(p) =
-e F_{1}[(p'-p)^{2}]\,(p' + p)^{\mu} \, \delta_{\mathfrak{s'},\mathfrak{s}}\,,
\label{4.1}
\end{eqnarray}
considering only the Dirac coupling.
From diagram~(b) of figure~\ref{fig:3}, we have
\begin{eqnarray}
{\cal M}^{(h \gamma)}_{pp \to pp \pi^{+} \pi^{-}} =
\bar{u}_{\mathfrak{s}_{2}}(p_{2}) 
\Gamma^{(\gamma pp)\,\mu}(p_{2},p_{b}) 
u_{\mathfrak{s}_{b}}(p_{b}) \frac{1}{t_{2}} 
{\cal M}_{\mu, \mathfrak{s}_{1}, \mathfrak{s}_{a}}(p_{3},p_{4},p_{1},q_{2},p_{a})|_{h}\,.
\label{3.6} 
\end{eqnarray}
Finally,
the $\gamma \gamma$-fusion contribution 
[figure~\ref{fig:3}~(c)]
is given by
\begin{align}
{\cal M}^{(\gamma \gamma)}_{pp \to pp \pi^{+} \pi^{-}} = \;&
\bar{u}_{\mathfrak{s}_{1}}(p_{1}) 
\Gamma^{(\gamma pp)\,\mu}(p_{1},p_{a}) 
u_{\mathfrak{s}_{a}}(p_{a}) \frac{1}{t_{1}}  \nonumber \\
&\times
\Big{[} 
{\cal M}^{(f_{2})}_{\mu, \mathfrak{s}_{2}, \mathfrak{s}_{b}}(p_{3},p_{4},p_{2},q_{1},p_{b})|_{\gamma} 
+
{\cal M}^{(\rm DS)}_{\mu, \mathfrak{s}_{2}, \mathfrak{s}_{b}}(p_{3},p_{4},p_{2},q_{1},p_{b})|_{\gamma}
\Big{]} \,.
\label{3.7} 
\end{align}

In the first approximation, we neglect absorption corrections 
due to the proton-proton interactions to the Born amplitudes
discussed above.
The absorption effect was discussed in Sect.~III~C of \cite{Lebiedowicz:2014bea}; 
it reduces the cross section for photoproduction processes 
by about 10\% at LHC energies.

\section{Results}
\label{sec-3}

In this section, we briefly discuss
the comparison of the tensor-pomeron model 
with the experimental data 
for the $\gamma p \to \pi^{+}\pi^{-} p$ reaction 
from \cite{H1:2020lzc}, and present predictions
for the $pp \to pp \pi^{+} \pi^{-}$ process.

\subsection{Our results for $\gamma p \to \pi^{+} \pi^{-} p$ compared with H1 data}
\label{sec-3}

\begin{figure}[h]
\centering
\includegraphics[width=6cm,clip]{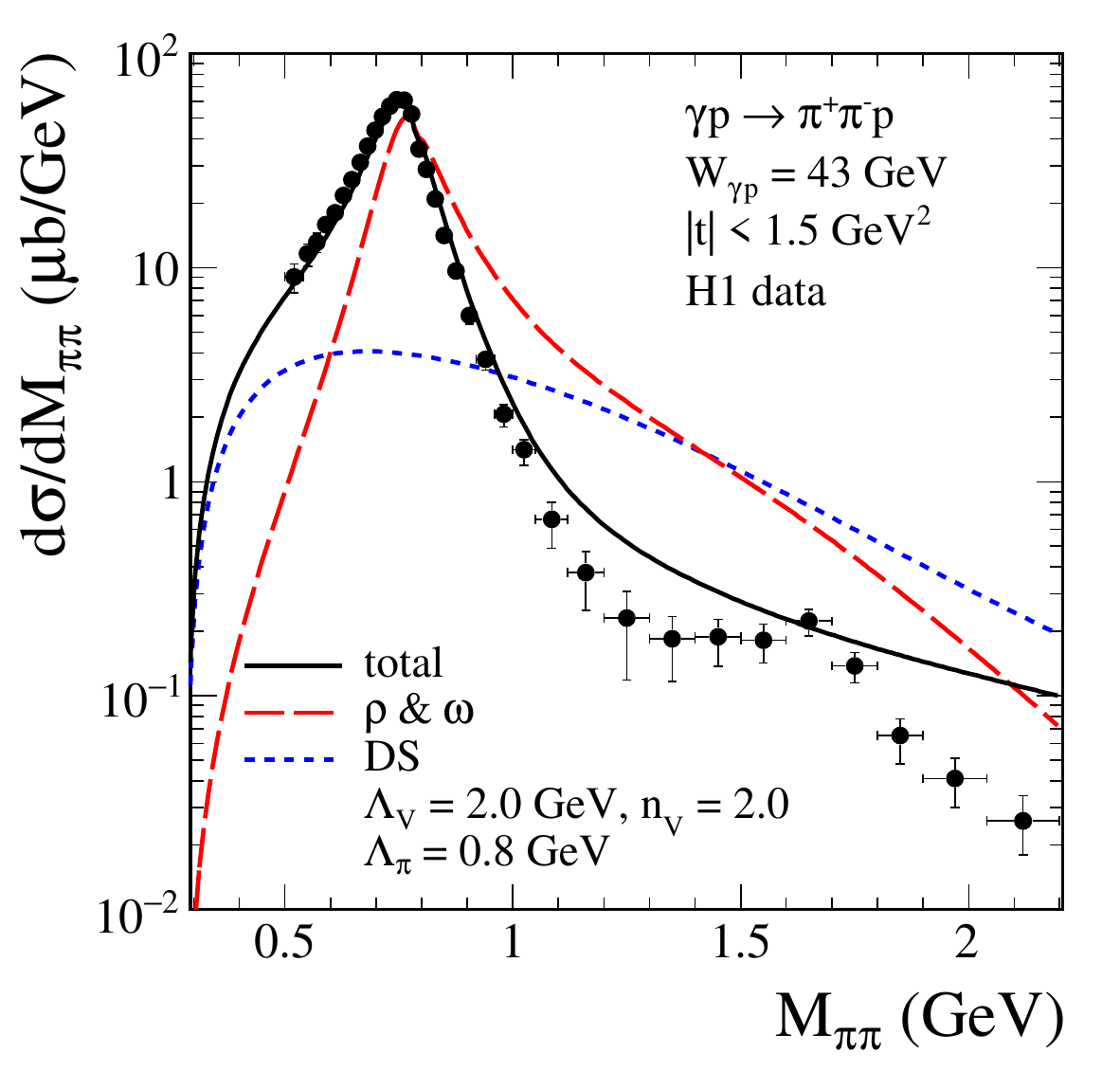}
\includegraphics[width=6cm,clip]{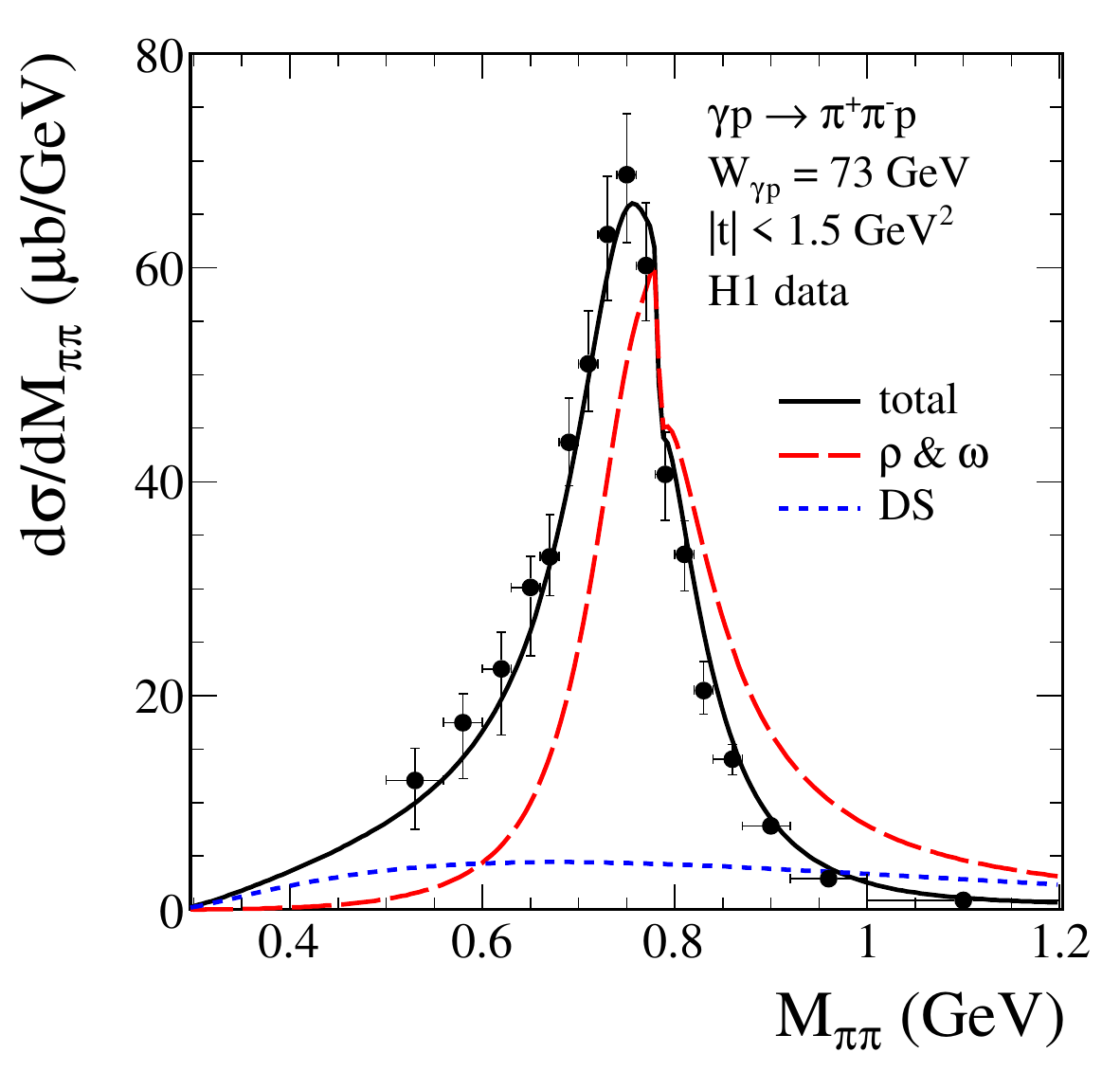}
\caption{
Comparison of our improved model \cite{Lebiedowicz:2026ixn},
including pion off-shell effects,
with the H1 experimental data from \cite{H1:2020lzc}.
Left panel: The invariant mass $M_{\pi\pi}$ distribution 
measured by the H1 Collaboration \cite{H1:2020lzc} 
for the kinematic range $20 < W_{\gamma p} < 80$~GeV 
($\langle W_{\gamma p} \rangle = 43$~GeV) 
and $|t| < 1.5$~GeV$^2$, compared to the model results evaluated
at fixed $W_{\gamma p} = 43$~GeV.
Presented are complete results (total) and results for the resonant ($\rho \,\& \,\omega$) and the non-resonant Drell-S\"oding (DS) contributions.
(From figure~2 of \cite{Lebiedowicz:2026ixn}).
Right panel:
Results for $W_{\gamma p} = 73$~GeV.
The meaning of the lines is the same as in the left panel.
(From figures~3 of \cite{Lebiedowicz:2026ixn}).}
\label{fig:4}
\end{figure}

In figure~\ref{fig:4}, we present our results for the two-pion invariant mass distributions for the $\gamma p \to \pi^{+} \pi^{-} p$ process.
Both resonant ($\rho(770)$ and $\omega$) and 
non-resonant Drell-S\"oding (DS) contributions are shown.
These results are compared with experimental data measured 
by the H1 Collaboration \cite{H1:2020lzc}.
The tensor-pomeron model with default parameters
gives a very satisfactory description of the H1 data 
for $M_{\pi \pi} < 1.2$~GeV.
The discrepancies that emerge at higher invariant masses, 
$M_{\pi \pi} > 1.2$~GeV, can probably be attributed to the 
contributions of higher-mass resonances, such as the 
$\rho(1450)$ and $\rho(1700)$ states, which are
not included in the present analysis.

In these calculations, we have included the pion off-shell effects
within the DS term. A comparison between the approaches 
with and without these effects, as well as further 
comparisons with H1 data \cite{H1:2020lzc} 
and also with ZEUS data \cite{ZEUS:1997rof}, 
can be found in \cite{Lebiedowicz:2026ixn}.

\subsection{Predictions for $pp \to pp \pi^{+} \pi^{-}$}
\label{sec-5}

\begin{figure}[h]
\centering
\includegraphics[width=6cm,clip]{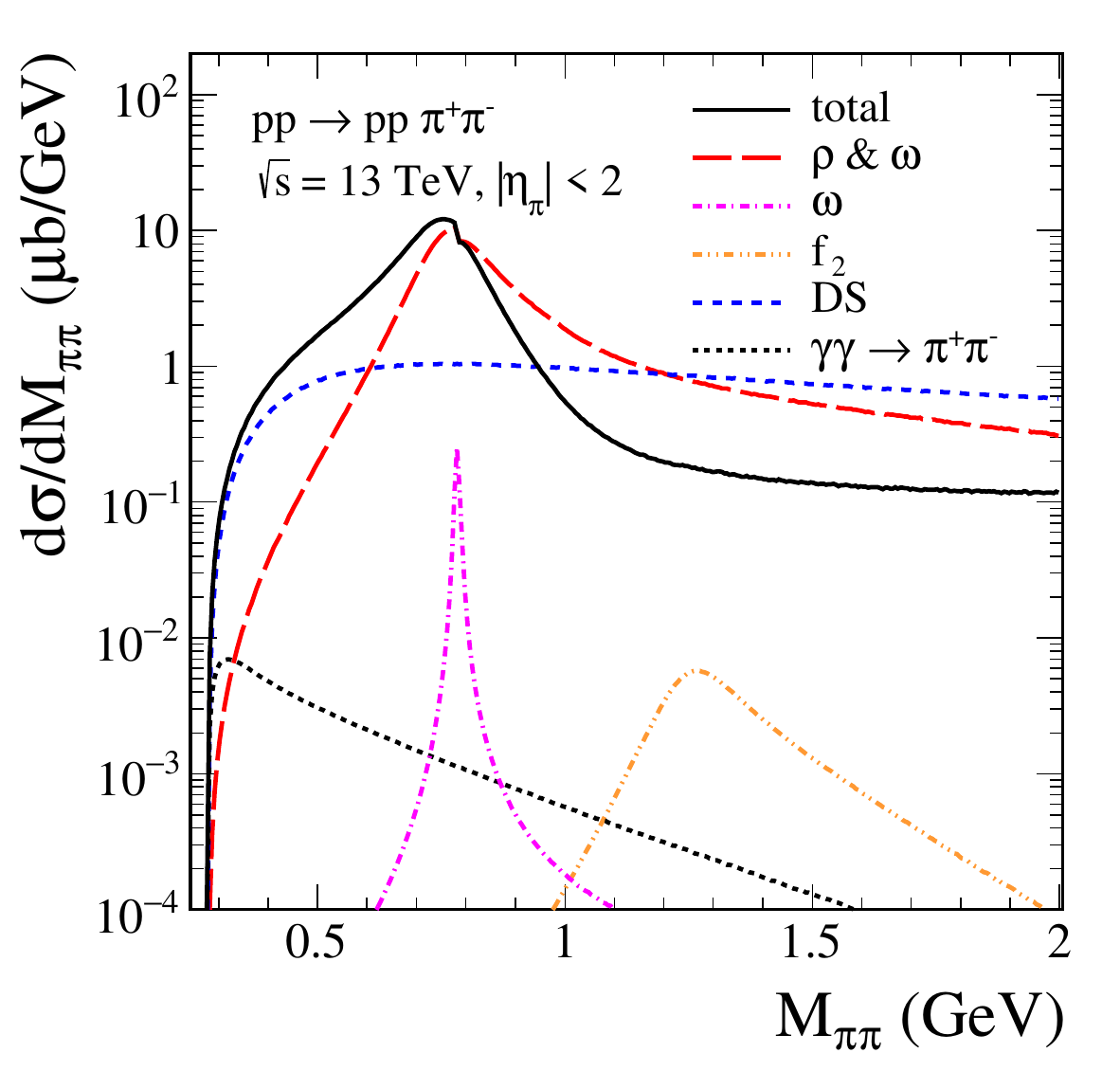}
\includegraphics[width=6cm,clip]{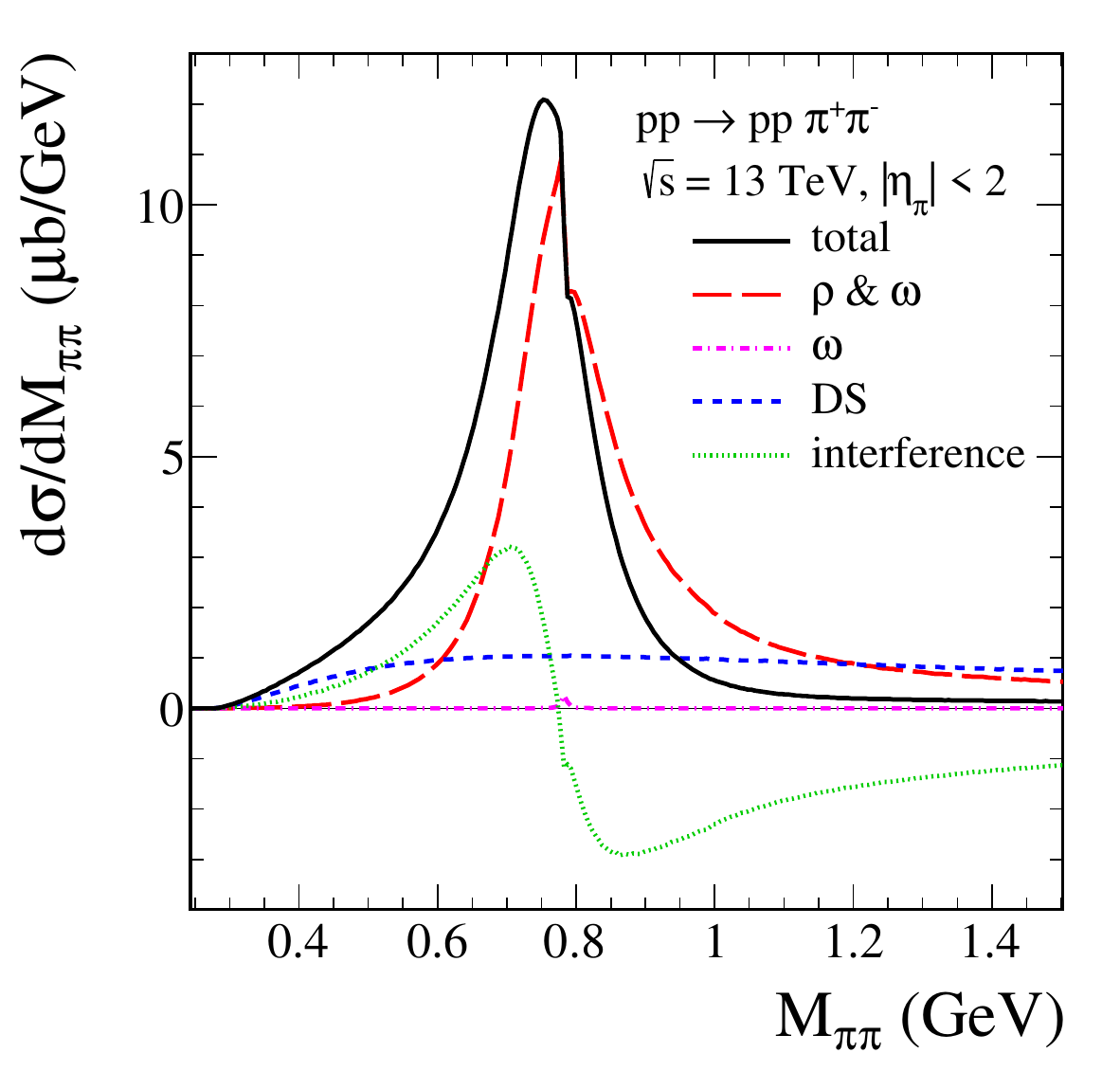}
\caption{
The two-pion invariant mass distributions for the $pp \to pp \pi^{+} \pi^{-}$ reaction calculated at $\sqrt{s} = 13$~TeV
and for $|\eta_{\pi}| < 2$.
In these calculations, 
we do not include the pion off-shell effects in the DS term.
Left panel: The full model (total) and individual contributions
from vector-meson production,
$f_{2}$ production, and non-resonant Drell-S\"oding (DS) processes are shown.
[From figure~9 (left panel) of \cite{Lebiedowicz:2025xob}].
Right panel: Detailed view of the main contributions on a linear scale.
The interference term between the resonant ($\rho$ and $\omega$) 
and DS contributions is represented by the green dotted line.
[From figure~7(b) of \cite{Lebiedowicz:2025xob}].}
\label{fig:5}
\end{figure}

Figure~\ref{fig:5} shows the differential cross sections as
a function of the two-pion invariant mass for the $pp \to pp \pi^{+} \pi^{-}$ reaction calculated at $\sqrt{s} = 13$~TeV 
and for pion pseudorapidity range $|\eta_{\pi}| < 2$.
The complete result (total) and the individual resonant and non-resonant (DS) contributions are presented.
For these predictions, in the DS term the pion off-shell effects 
are not taken into account.
As shown in the left panel of figure~\ref{fig:5}, 
the production of the $f_{2}(1270)$ resonance
is more than three orders of magnitude smaller than
the production of $\rho(770)$.
Furthermore, the $\gamma \gamma \to \pi^{+} \pi^{-}$ fusion 
contribution is found to be negligibly small in this kinematic regime.
In the right panel of figure~\ref{fig:5},
the skewing of the $\rho(770)$ line shape caused 
by the interference of the $\rho$ and DS contributions 
is clearly visible. In addition, the characteristic
$\rho$-$\omega$ interference pattern is prominently featured.

\newpage

\section{Conclusions}
\label{sec-6}

Our main points are as follows.


We have reported results from
\cite{Lebiedowicz:2025xob,Lebiedowicz:2026ixn} concerning
a new calculation of the non-resonant Drell-S\"oding (DS)
term for the $\gamma^{(*)} p \to \pi^+ \pi^- p$ reaction. 
This new result is derived within the framework of Quantum Field Theory (QFT).
We have extended the calculations for the $\gamma p \to \pi^+ \pi^- p$
process, which were originally presented for real photons in \cite{Bolz:2014mya}, 
to include virtual photons in the range $0 < Q^{2} = - q^{2} \lesssim 0.5$~GeV$^2$.

We have shown that a better treatment of the Drell-S\"oding contribution than that presented in \cite{Bolz:2014mya} is needed to describe 
the H1 data \cite{H1:2020lzc}, specifically in
the region of low di-pion invariant masses and in describing 
the skewing of the $\rho(770)$ spectral shape.
Detailed comparisons of our improved model
for the $\gamma p \to \pi^+ \pi^- p$ reaction with current HERA data,
and in the future with measurements from the EIC and LHeC,
will offer a valuable opportunity to validate our model
and to constrain its (few) parameters.
The latter are mainly related to the shapes of some form factors.

Our results will have a substantial impact on experimental hadron physics. 
Currently, the central exclusive production (CEP) of $\pi^{+} \pi^{-}$ 
pairs is being actively investigated in the $pp \to pp \pi^{+} \pi^{-}$ 
process at the LHC. Moreover, this formalism can be directly used 
for the photoproduction of pion pairs in ultraperipheral $p$A and AA 
collisions at the LHC. In the future, the $\gamma^{(*)} p \to \pi^+ \pi^- p$ 
reaction will be a key topic for the experiments at next-generation 
electron-proton/ion colliders (EIC, LHeC). 
For all these processes, we provide 
a robust and reliable theoretical framework firmly grounded in QFT.


%
%

\end{document}